# Electrical and structural properties of pure and dysprosium doped $Na_{0.5}Bi_{0.5}TiO_3$ system: DFT and Monte Carlo simulation

Manal Benyoussef[1], Halima Zaari[2], Jamal Belhadi[1], Abdelilah Lahmar[1], Youssef El Amraoui[2,3], Hamid Ez-Zahraouy[2], Mimoun El Marssi[1]

[1] Laboratory of Physics of Condensed Matter (LPMC), University of Picardie Jules Verne, Scientific Pole, 80039 Amiens Cedex 1, France

[2] Laboratory of Condensed Matter and Indisciplinary Science (LaMCScI), Faculty of Science, Mohammed V University, Rabat, Morocco

[3] National School of Arts and Crafts, Moulay Ismail University, Meknes, Morocco

**Abstract**

The chemical ordering, electrical, optical, and magnetic properties of $Na_{0.5}Bi_{0.5}TiO_3$ (NBT) and 25% dysprosium doped NBT (DyNBT) were investigated in the framework of first-principles calculations using the full potential linearized augmented plane wave (FP-LAPW) method based on spin-polarized density functional theory implemented in the WIEN2k code. We demonstrated that NBT structure is stable in the 001 A-site configuration, while DyNBT presents an A-site disorder perceived by the minimal energy difference between the different A-site configurations. A significant magnetic moment of $5\mu_B$ emerges in DyNBT system, while NBT is known to be non-magnetic. Dysprosium in NBT matrix seems to form an ionic bonding with oxygen atoms whereas Bi-O forms covalent bonding which is responsible for the decrease of the polarization value from 42.3 µC/cm² for NBT to 22.08 µC/cm² for the doped compound. In the second part, the transition temperature and the hysteresis loops of $Na_{0.5}(Bi_{1-x}Dy_x)_{0.5}TiO_3$ system (x = 0 – 25%) were investigated using the Monte Carlo simulation. We observed a decrease in the transition temperature as a function of dysprosium introduction. We pointed out from the hysteresis loops, an apparent decrease of the coercive field together with the remanent polarization as a function of doping and also as a function of temperature. Our proposed model was seen to approach the values of experimental studies.

Corresponding author: manalbenyoussef@gmail.com

# 1. Introduction

Since the European Union has restricted the use of lead in electronic equipment (2011/65/Eu (RoHS) [1], due to the concern about human health and the environment, research activity got oriented into lead-free materials with enhanced properties. Among the most studied lead-free ferroelectric perovskite having promising properties, one can found the A-site mixed relaxor ferroelectric $Na_{0.5}Bi_{0.5}TiO_3$ (NBT) system. NBT system has been extensively studied by both experimental and theoretical methods due to its high piezoelectric response, high spontaneous polarization, and rich variety of dielectric behaviors [2–8]. The latter material was thence seriously considered to replace lead-based materials in power electronic applications [9]. The pure NBT system is known to undergo several phase transitions (PTs). Going through cubic, tetragonal, and rhombohedral structures [10]. In the high-temperature range (>790K), NBT has a paraelectric cubic structure (Pm3-m) with no octahedral tilting ($a^0a^0a^0$). The tetragonal structure (640K–790K) has an in-phase $TiO_6$ octahedral tilting ($a^0a^0c^+$), which is weakly polar due to unequal antiparallel A and B cationic displacement along the c axis [11]. The non-centrosymmetric rhombohedral (R3c) structure stabilizes at room temperature (RT) with an antiphase $TiO_6$ octahedral tilting according to the modified glazer notation ($a^-a^-a^-$). The spontaneous polarization at room temperature result from cationic displacements along the [111] direction, with a high remanent polarization value ($P_r$ =38 µC/cm²). Nevertheless, the large coercive field ($E_c$ =73 kV/cm), the high conductivity and the high dielectric losses of the sodium bismuth titanate makes researchers looking for new NBT-based systems which can have improved ferroelectric and piezoelectric properties. Several studies got interested in binary, and ternary systems based on NBT [12–14]. Besides, doping with rare earth elements has been reported to be an excellent alternative to improve the properties of pure NBT [4,5,15–19]. For instance, we demonstrated in our previous experimental work, an enhanced resistivity as well as improved energy storage properties at high temperatures in NBT doped dysprosium element [5]. Based on previous experiments on lead-based ferroelectric systems, doping with rare-earth elements can induce a change in the ordering degree of A/B-site cations, and can also induce random fields/bonds, which gives rise to local structural heterogeneity that are important in the view of improving the physical properties of materials [20–22]. For instance, Rare earth doping has been reported to considerably improve the piezoelectric response of the lead-based relaxor ferroelectric PMN-PT material due to the induced nanoscale structural heterogeneities [23]. This strategy has been recently employed in the lead-free antiferroelectric $AgNbO_3$ system to improve the energy storage density by the use of Samarium ($Sm^{3+}$) dopant in order to decreases the dielectric loss and increase the critical antiferroelectric–ferroelectric phase transition electric field [24]. Thence, it would be interesting to observe theoretically what would be the effect of rare-earth doping on the physical properties of NBT system, and especially on the A/B-site ordering of the perovskite structure.

Although several studies were interested in the investigation of the A-site chemical ordering of the pure NBT system by both theoretical and experimental methods, this topic remains much debated. Usually ordering or disordering of the AA′ cations of the AA′BO$_3$ perovskite depends on the ionic radii sizes, bonding preferences, and oxidation states of cations [25,26]. Cationic ordering is generally driven by electrostatic considerations due to different ionic radii and oxidation states of cations, whereas cationic disordering is driven by configurational entropy when cations have similar ionic radii and oxidations states. In addition, the charge difference ($\Delta q$) value is essential for determining either order ($\Delta q > 2$) or disorder ($\Delta q < 2$) in cations is favored. In the case of $\Delta q = 2$, the arrangement is not resolved and can be fully ordered/disordered or partially ordered [27]. In NBT system, $Na^+$ and $Bi^{3+}$ have a charge difference equal exactly to two ($\Delta q = 2$), in addition to similar ionic radii ($r_{Na+} = 1.02$ Å, $r_{Bi3+} = 1.03$ Å). Thence, full order/disorder, or partial order may be favored in the A-site of NBT. Using single-crystal XRD measurements, Park *et al.* reported on a low degree of ordering in the cubic phase of NBT system. This last assumption was later confirmed by Dorcet et Tortillard based on TEM images and electron diffraction. Besides, using Raman spectroscopy, Petzelt *et al.* described the occurrence of an A-site chemical ordering [28]. On the other hand, no indication of chemical ordering has been revealed using high angle dark field (HAADF) scanning transmission electron microscopy (STEM) images. Theoretical works were also interested in the stability of different A-site occupations in NBT system looking for the most stable configuration, i.e., having the lowest energy [29–33]. The focus is generally put on the study of the pure NBT compound, in its high-temperature cubic structure. Thence, the first aim of the present work is to investigate in detail the chemical ordering of the pure and dysprosium doped NBT system in their rhombohedral (R3c) structure. Based on our knowledge, there is no previous theoretical study about the chemical ordering of rare-earth doped NBT system using the DFT method. Furthermore, this work will give interesting results about the electronic, optical, electrical, and magnetic properties of both NBT and NBT doped systems by the use of ab initio method in the framework of the Generalized Gradient Approximation (GGA) implemented in the WIEN2k package.

Moreover, Monte Carlo simulation is known to be a powerful method for the investigation or prediction of magnetic and/or ferroelectric properties of compounds [34–37]. Using an effective Hamiltonian, Zhong *et al.* investigated by Monte Carlo simulation the phase sequence, transitions temperatures, spontaneous polarization, and latent heats of BaTiO$_3$ (BTO) compound [36]. Further studies are found in the literature survey on the simulation of the ferroelectric properties of compounds. For example, Bedoya-Hincapié *et al.* developed a Diffour Hamiltonian taking into account the dipole-dipole interaction in order to study the ferroelectric response of bismuth titanate Bi$_4$Ti$_3$O$_{12}$ thin film [37]. They presented the effect of stress and temperature application on the hysteresis loops. The simulation of the ferroelectric properties of the lead-based Pb(Zr$_x$Ti$_{1-x}$)O3 PZT system has also been widely studied using Monte Carlo simulations in bulk and thin-film forms [38–41]. For instance, a Janssen-like Hamiltonian taking into account the magnetoelectric interactions was used in order to simulate the ferroelectric behavior of the ferroelectric (PZT)/ferromagnetic (LSMO) bilayer [40]. As far as we know, the ferroelectric behavior of the A-site mixed sodium bismuth titanate material has not yet been investigated by Monte Carlo simulation. For that reason, we will present in a second part a detailed Monte Carlo simulation on Na$_{0.5}$(Bi$_{1-x}$Dy$_x$)$_{0.5}$TiO$_3$ (xDyNBT) system with x = 0 – 25%. Therefore, a structurally based ferroelectric model will be simulated using a Monte Carlo simulation in the case of the rhombohedral perovskite structure. We will focus on the simulation of the electrical properties of xDyNBT systems. The phase transition temperature will be extracted for different Dy$^{3+}$ concentrations. Electrical hysteresis loops will be simulated in order to investigate the ferroelectric response to an applied electric field. The impact of the temperature on the polarization and the coercive field will also be discussed. To the best of our knowledge, this study constitutes the first simulation of the electrical properties of NBT based materials using a ferroelectric model.

## 2. Computational methodology

All calculations were performed using the full-potential linearized augmented plane wave (FP-LAPW) method based on spin-polarized density functional theory, as implemented in the WIEN2k code [42]. The muffin tin radii (RMT) were taken to be 2.32 a.u. (atomic units) for Na, and 2.38, 2.38, 1.84, 1.66 a.u. for Bi, Dy, Ti, and O, respectively. The exchange-correlation energy is approximated within the generalized gradient approximation (GGA). We use 200 k-points for Brillouin zone integration [43]. The rhombohedral structure of NBT with R3c ($C_{3v}^6$) space group was taken in this study due to its ferroelectric properties. A relaxation of the atomic positions was performed at the experimental cell volume (a=5.505 Å, α=59.78°) [5] of the unit-cell structure of NBT (figure 1 (a)). Relaxations of supercell structures have been made and no change in the properties was observed. Table 1 groups the theoretical and experimental lattice constant together with the angle (α). The optical properties were obtained from the complex dielectric function ε(ω) which is given by:

$$\varepsilon(\omega) = \varepsilon'(\omega) + i\varepsilon''(\omega) \qquad (1)$$

Where ε' (ω) and ε'' (ω) are the real and imaginary parts of the dielectric function. From ε(ω), all other optical properties can be extracted, as the absorption coefficient α(ω) by the following expression.

$$\alpha(\omega) = \sqrt{2}\omega[\sqrt{\varepsilon'^2(\omega) + j\varepsilon''^2(\omega)} - \varepsilon'(\omega)]^{1/2} \qquad (2)$$

In order to investigate the chemical ordering in the A-site, a supercell of 2×1×2 (40 atoms) was considered. The supercell contains eight A-cations (4Na and 4Bi) on the rhombohedral structure. The investigated supercell was only considered in order to substitute 1/4 bismuth atom by one dysprosium atom, which will give a substitution percent of 25%$Dy^{3+}$. A bigger supercell (80 atoms) may be considered in order to substitute 1/8 bismuth atom by dysprosium resulting in 12.5%$Dy^{3+}$. Therefore, we studied three different A-site occupations: the 111, 110, and 001 configurations, as depicted in figure 1. These last configurations were made by changing the emplacement of atoms in order to obtain the wanted configuration. Berry phase PI approach [44] implemented in the Wien2k code permits us to investigate the spontaneous polarization of our studied configurations.

**Table 1.** Theoretical and experimental lattice constant $a_0$ (Å), and angle α (°) of R3c NBT.

|  | **Theory** | **Experiment** [5] | **Theory** [10] | **Theory** [26] |
|---|---|---|---|---|
| $a_0$ | 5.550 | 5.505 | 5.5051 | 5.421 |
| α | 59.68 | 59.78 | 59.8027 | 59.499 |

### 3. Results and discussion
### 3.1. Ab initio calculations
*3.1.1  A-site Ordering and Relative Stability*

As mentioned before, NBT crystallizes in the rhombohedral structure with R3c space group symmetry at room temperature. Figure 1 (b-d) groups supercells with different A-site ordering, performed in order to investigate the chemical ordering as well as its effect on the physical properties of NBT. Thus, we consider 2×1×2 supercells containing 40 atoms consisting of four Bi, four Na atoms in the A-site, and eight Ti atoms in the B site of the ABO$_3$ perovskite, in addition to twenty-four O atoms. We were especially interested in three A-site ordering, known as the layered (001), columnar (110), and rock-salt (111) configurations.

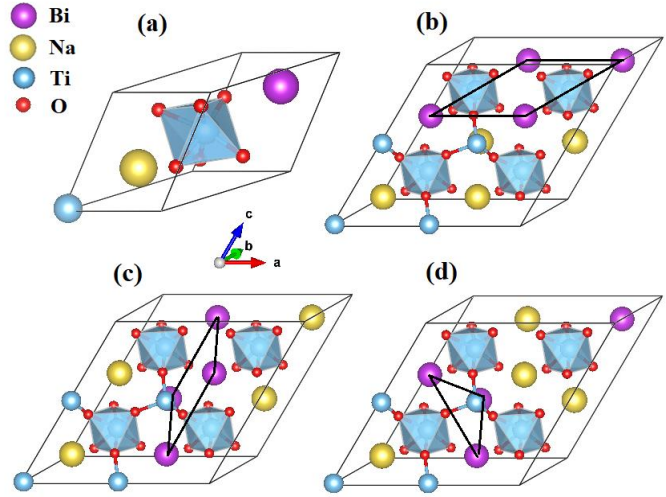

**Figure 1:** (a) Elementary cell and supercells (2×1×2) having (b) 001, (c) 110, and (d) 111 A-site occupations of pure NBT.

These A-site arrangements differ in the way the AO$_3$ or A′O$_3$ octahedra connect in the AA′BO$_3$ perovskite (A=Na, A′=Bi, B=Ti). Therefore, the layered arrangement allows for the connectivity between the A′O$_3$ octahedra in two dimensions (2D), whereas the columnar arrangement allows only one dimension of connectivity between the A′O$_3$ octahedra (1D). In the case of the rock-salt configuration, there is no connectivity between the A′O$_3$ octahedra in the structure (0D). Generally, B-site ordering is more common than the A-site ordering, which is due to the difference in the oxidation states (Δq) [45]. In the case of B/B′-site Δq may reach seven, whereas in the A/A′-site Δq is limited to two or less. In addition, the anion environment differs depending on the A-site arrangement. In the layered arrangement, there are three different anion environments. One where the anion is surrounded by four Na, a second where the anion is surrounded by four Bi, and a third where the anion is surrounded by two Na and two Bi in a cis configuration [46]. The last anion environment is the majoritarian one for the layered arrangement. On the contrary, the rock-salt configuration has only one anion environment, in which the anion is coordinated by two Na and two Bi cations in a trans configuration. Regarding the columnar arrangement, there are three anion environments: two-third having two Bi and two Na in a cis configuration and one third having two Bi and two Na in a trans configuration [31].

In order to see the impact of the different A-site configurations on the stability of the structure, we compared the energy of the different structures by considering a relative energy ($\Delta E_{relative}$) where the 001 direction is the reference ($\Delta E_{relative}$ is obtained by subtracting the energy of the 001 configuration for all compositions) as shown in Table 2. Therefore, by comparing the relative obtained energies, we can conclude that the favored structure i.e., having the lowest relative energy, is the layered arrangement, followed by the rock-salt 111 and finally by columnar 110 A-site order. Generally, the layered order is favored by the A-site, whereas the rock-salt order is favored by the B-site which is due to the fact that the 111 B-site ordering maximizes the separation between the highly charged B′ cations allowing the stabilization of the structure [45]. The stabilization of the layered arrangement may be caused by a single environment for all anions. In this last, anions sit in a site with inversion symmetry and thus the Na-O and Bi-O bonds length remain similar. We agree that the energy difference between each configuration is noticeable of at least 330 meV/f.u, which expresses a 001-chemical ordering in NBT matrix. Gröting *et al.* also reported higher stability of the 001-configuration due to the hybridization of Bi 6p and O 2p state making $Bi^{3+}$ lone pairs stereochemically active [31]. Contrary to Burton and Cockayne that reported lower energy for the "crisscross" arrangement of Na/Bi ions [32].

**Table 2:** Relative energy ($\Delta E_{relative}$) per formula unit, electronic band gap, Berry phase polarization (P), and magnetic moment of the different A-site arrangements in NBT and DyNBT systems.

| Compound – A site configuration | $\Delta E_{relative}$ (meV/formula unit) | Band Gap (eV) | $P_s$ ($\mu C/cm^2$) | M ($\mu_B$) |
|---|---|---|---|---|
| **NBT – 110** | 440 | 2.73 | 45.9 | 0 |
| **NBT – 111** | 330 | 2.23 | 43.6 | 0 |
| **NBT – 001** | 0.00 | 1.67 | 42.3 | 0 |
| **DyNBT – 110** | 33 | 1.83 | 29.70 | 5.04 |
| **DyNBT – 111** | 22 | 2.24 | 33.03 | 4.96 |
| **DyNBT – 001** | 0.00 | 1.91 | 22.08 | 4.99 |

The investigation of rare-earth doping on the chemical ordering of NBT matrix was done by the study of dysprosium ($Dy^{3+}$) introduction in NBT. For that purpose, we substitute one Bi atom by one Dy atom, which is equivalent to 25%$Dy^{3+}$ doping NBT in the three earlier A-site arrangement (R3c structure); layered, columnar, and rock-salt configurations. Therefore, we compute here too the relative energies per 001 – DyNBT formula unit presented in Table 2. The lowest energy is given for DyNBT in the layered arrangement with a maximum energy difference of 33 meV/f.u between configurations. The lower ionic radii of dysprosium ($r_{Dy3+}$ = 0.91 Å) compared to that of the bismuth will increase the size difference between A and A′ cations in the AA′BO3 perovskite in certain regions favorizing an A-site ordering while in other regions the similar ionic radii of Na and Bi will favorize an A-site disorder. However, the substitution of bismuth by dysprosium atoms, in a random way, introduce a quenched disorder. Therefore, the weak computed relative energy (per NBT-001 formula unit) difference may express disordered or partially ordered regions in the DyNBT system. In addition, the introduction of dysprosium seemed to decrease the spontaneous polarization of the system from ~45 µC/cm² for NBT to ~30 µC/cm² for the DyNBT system. Moreover, dysprosium is also known for its magnetic properties driven by the partially filled f-orbitals (4f[10]). Interest has been devoted to the introduction of magnetic ions in ferroelectric systems, in order to create novel multiferroic systems. As observed in Table 2, a magnetic moment of about 5 Bohr magneton appears in the DyNBT system in contrary to the non-magnetic NBT system expressed by the zero computed magnetic moment. Thereby, the coexisting magnetic and electric polarization may allow an additional degree of freedom in the design of novel devices (transducers, actuators, storage devices, etc.). In addition to multiple-state memory elements, where data can be stored either by magnetic and electric polarization [47].

*3.1.2 Electronic and optical properties*

The total and partial density of electronic states of the three A-site arrangements of the NBT system are given in figure 2 (a-c). As we can see, the Valence Band (VB) is essentially constituted from oxygen (O) 2*p* orbitals with titanate 3*d* and bismuth 6*p* orbitals contribution, in addition to a sharp contribution of Bi 6*s* states that lies deep below the VB (not shown). Thus, a strong hybridization occurs between Bi 6*p*, Ti 3*d* orbitals and O 2*p* orbitals in the VB. Regarding the low energy values of the CB (Conduction Band), the Ti 3*d*, together with Bi 6*p*, are the dominating states. There are three significant differences between the different A-site ordering on the DOS. First, the bandgap is dependent on the orientation. Thus, the conduction band is shifting to higher energies by ~ 1 eV for 110 and by ~ 0.5 eV for the 111 in comparison to the 001 arrangement. Second, the oxygen 2p states near the maximum valence band (VBM) has a different response, which evidences the different anion environment for the different A-site arrangement. The third difference in the DOS is occurring for Ti 3d states. We observe that for the two 001 and 110 configurations sharper peaks are appearing, which is due to the splitting of d orbitals. Remind that the orbital d is known to split in the octahedral crystal field into $t_{2g}$ and $e_g$ states which split again into bonding and anti-bonding states, which are due to the interaction with oxygen 2p states. The electronic , of the three configurations, is comparable with earlier studies of NBT [48,49]. Gröting *et al.* reported an increase of the bandgap for structures having local ionic displacements [31].

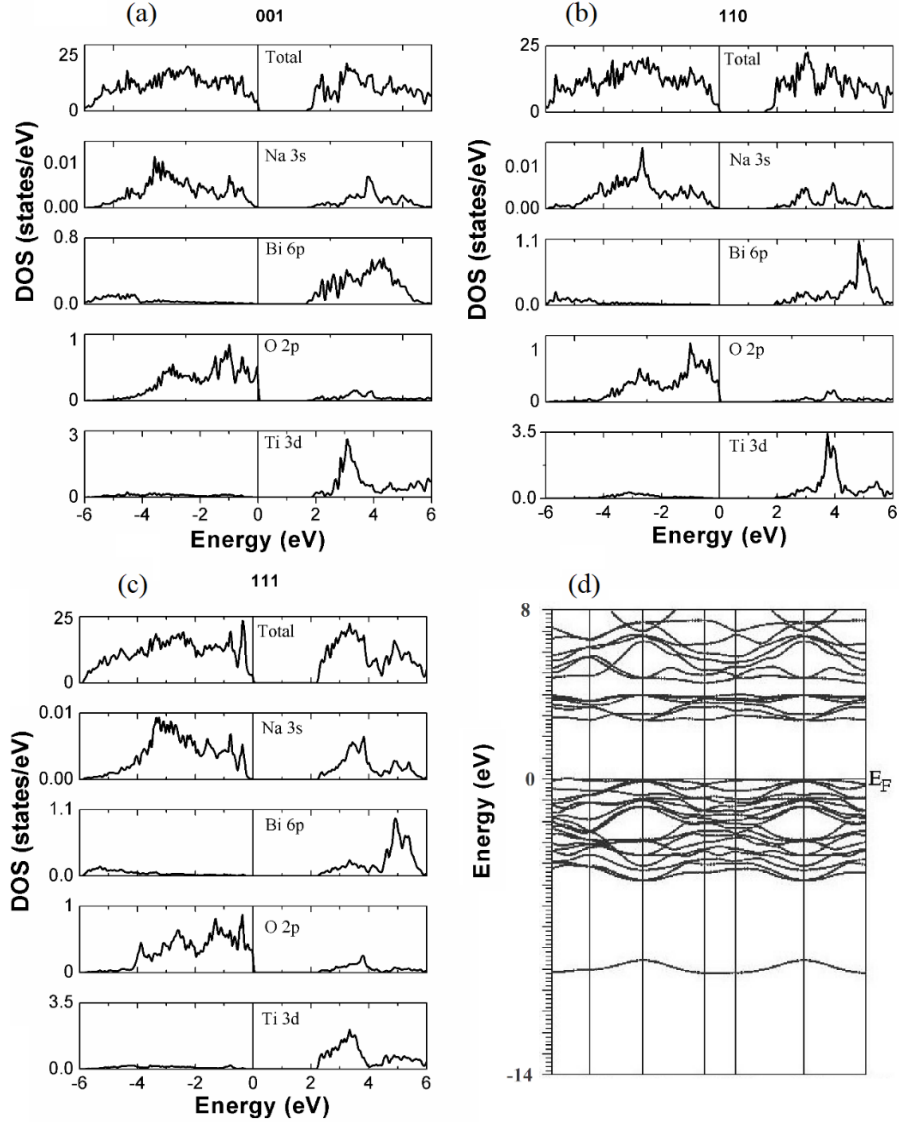

**Figure 2:** Total and partial density of states versus the energy of (a) 001, (b) 110, and (c) 111 A-site ordering of NBT. (d) The band structure of 001-NBT.

The band structure presented in figure 2 (d) confirms the electronic bandgap value and permits us to investigate its nature. It is clearly observed that the top of the VB and the bottom of the CB are both situated in the highly symmetric Γ point of the Brillouin zone, which expresses a direct bandgap process for all configurations. Figure 3 presents the charge density distribution of the layered configuration of NBT. This mapping indicates well the strong Bi-O and Ti-O hybridization. These strong interactions are pertinence in the emergence of the ferroelectricity in A and/or B site. Moreover, we can also conclude from figure 3 about the nature of interactions of the different atoms. In fact, bismuth and titanate atoms are seen to share covalent bonding with oxygen atoms. On the other side, sodium prefers ionic bonding with oxygen.

The dipole interaction values were computed using the energy and polarization values obtained, by GGA approximations in DFT framework, for different configurations of dipoles, by moving the cations which induce ferroelectricity in the system from their initial positions. The values 0.61 eV C/m², 0.29 eV C/m², and 0.066 eV C/m² are respectively given for the dipole interaction values of Na/Bi, Na/Na, and Bi/Bi.

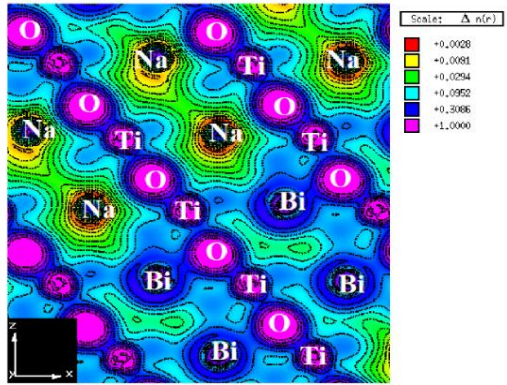

**Figure 3:** The Charge density distributions of NBT-001 orientations (supercell 2×1×2).

We were also interested in the effect of rare earth introduction on the electronic properties of the NBT system. Figure 4 (a-c) presents the density of electronic states of DyNBT in the three A-site configurations. Remind that rare earth elements are characterized by the partially filled 4f orbitals, which split into three-part due to the tetrahedral field: the triply degenerated $t_{1g}$ and $t_{2g}$, in addition to the singly degenerated $a_{2g}$. Notice that the electronic configuration of $Dy^{3+}$ is $4f^9$. Therefore, the 4f electrons will occupy totally the majority spin states. Afterward, they will occupy two-thirds of the minority $t_{1g}$ spin states which have the lowest energy. As observed from the partial density of electronic states of dysprosium element given in figure 4 (a-c), the magnetic moment was mainly obtained by the RE 4f orbital. An observable hybridization between O 2p and Dy 4f orbitals is evidenced by the DOS. The effect of the different A-site configurations remains similar to the NBT case. However, 6s orbital of bismuth is observed to sharpen for the layered arrangement. Sharper is Bi 6s orbital peak weaker will be the Bi-O interaction and thus, a lower dipole moment will be obtained which explains the lower polarization value of the 001-DyNBT configuration compared to the two other A-site arrangements (Table 2).

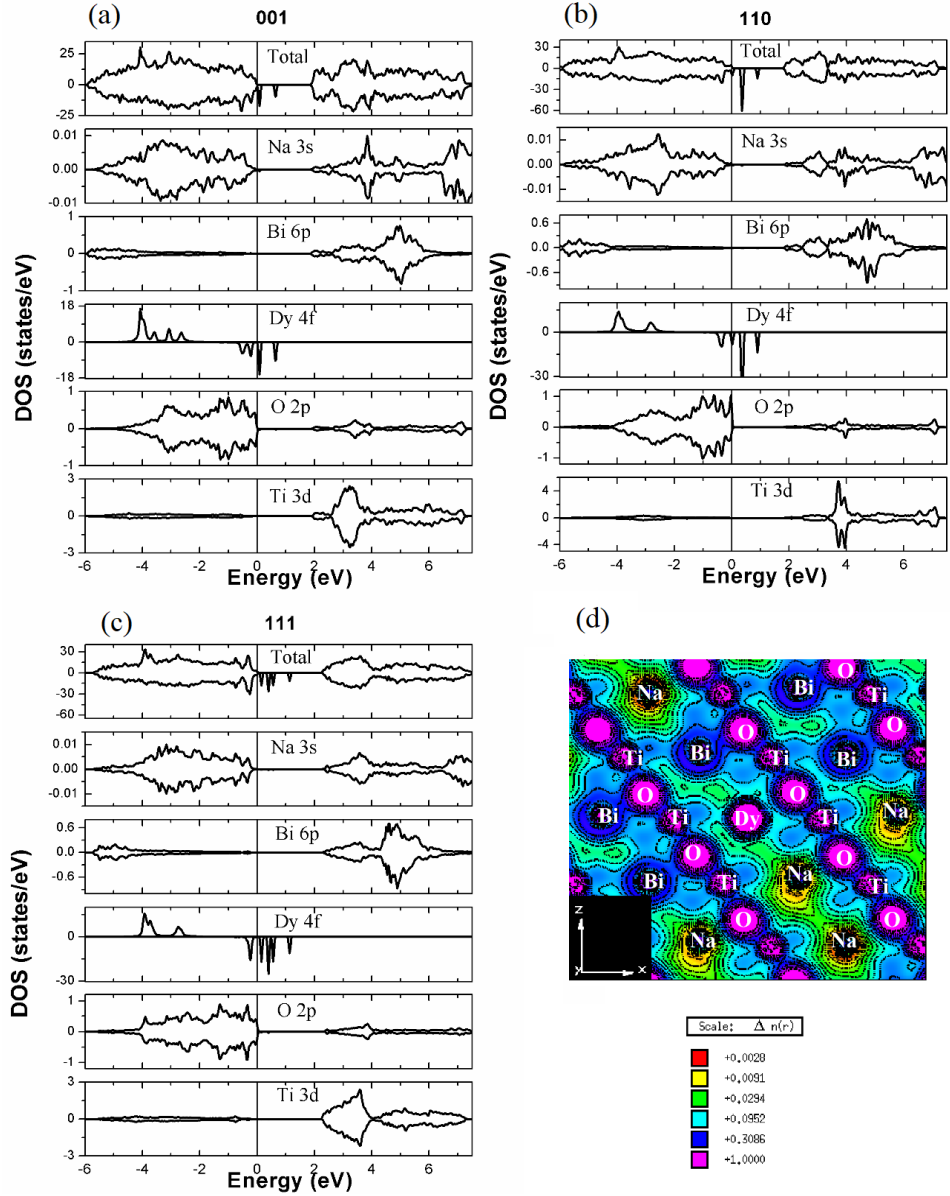

**Figure 4:** Total and partial density of states versus the energy of (a) 001, (b) 110, and (c) 111 A-site ordering **(d)** charge density distributions of DyNBT system (001 A-site ordering).

The other observable difference in the DOS is seen in the PDOS of dysprosium were depending on the A-site arrangement different 4f orbital contributions are observed. For the layered arrangement, we observe four peaks in the spin-up channel of the valence band from -4.45 eV to -2.23 eV. Concerning the columnar arrangement, only two peaks are rising from -4.37 eV to -2.32 eV in the spin-up channel of the valence band. On the contrary, the spin-up channel of the valence band in the rock-salt arrangement contains three peaks from -4.28 eV to -2.27 eV. The spin-down channel in the valence band also differs with two, one and a half, and one peak is observed for the 001, 110, and 111 A-site arrangement, respectively. We conclude that the A-site ordering does impact the magnetic moment as well as the polarization moment of the doped system. Figure 4 (d) presents the charge density distributions of the 001 - DyNBT system. First, we can notice that Bi-O and Dy-O have different interaction types. Dysprosium is seen to form ionic bonding with the oxygen instead of the covalent bonding of Bi-O. The different bonding types can influence the ferroelectric properties of the system. In fact, a reverse Monte Carlo simulation on neutron total elastic scattering of NBT system demonstrated an affinity of bismuth toward off-centering which is due to the Bi-O covalent bonding [50]. The off-centering of bismuth from the surrounding oxygen anions is one of the phenomena responsible for driving ferroelectricity in NBT system. Therefore, replacing the covalent bonding of Bi-O by the ionic bonding of Dy-O will weaken the polarity in DyNBT system. This is in good agreement with the reduction of the polarization values from 42.3 µC/cm² for NBT to 22.08 µC/cm² for 25DyNBT in the 001 A-site configuration (see table 1).

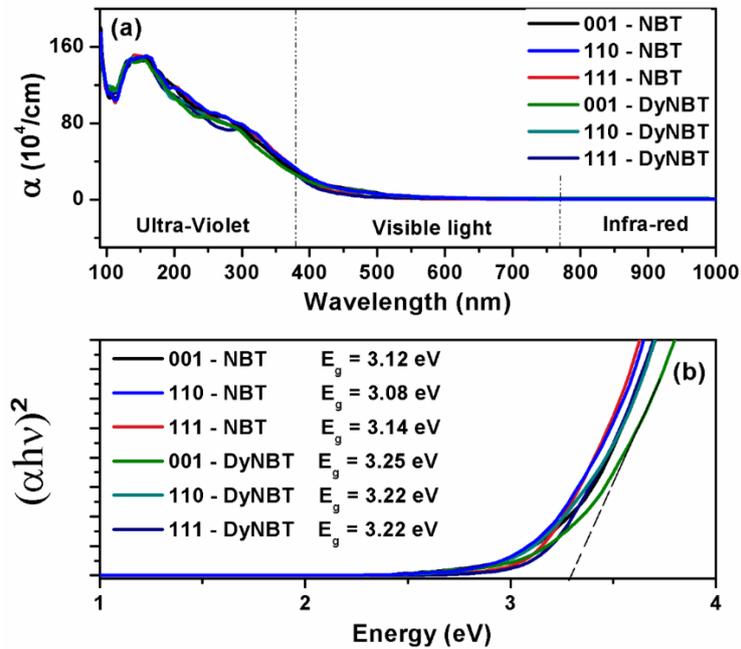

**Figure 5:** **(a)** The absorption as a function of wavelength, **(b)** the optical bandgap as a function of energy for NBT and DyNBT for different A-site orientations.

In order to confirm the change variation in the electronic band gap, we investigate the optical properties of NBT and dysprosium doped NBT. Thus, figure 5 (a) presents the absorption of our systems (pure and doped) with the different A-site ordering considered in this study. All studied systems seem to have the same absorption behavior. Still, a small difference between the different A-site ordering in the visible region can be observed. The maximum absorption value is observed at the lower wavelength (higher energies) and is seeming to decrease for higher wavelength (lower energies), to reach zero near 790 nm. Therefore, intense absorption in the ultraviolet region, with a non-negligible absorption in the visible light region is observed.

Good transparency in the infra-red region is also highlighted. Our systems appear to be potential for optoelectronic UV applications. With the use of the Tauc relation (Eq. (3)), we could examine the optical bandgap nature and value of the studied systems.

$$(\alpha h\nu)^n = A(h\nu - E_g), \quad (3)$$

Where the value of the parameter (***n***) can take different values depending on the transition process, i.e., ***n*** = 2 for a direct bandgap, and ***n*** = 1/2 for an indirect bandgap. **E$_g$** is the optical band gap value, **h$\nu$** is the photon energy, and (**A**) is a constant. The optical band gap values of NBT and DyNBT systems for different A-site ordering were calculated by extrapolating the linear portion of the curve to E-axis (figure 5 (b)). The values of the optical band gap of NBT (001, 110 and 111) are in good agreement with experimental results obtained in ref.[51]. Doping NBT with Dy$^{3+}$ seems to increase E$_g$ for all configurations, from ~3.10 eV for NBT to ~3.20 eV for DyNBT. An increase of the bandgap value with rare earth introduction was also reported in ref.[52] for Nd$^{3+}$ doped NBT.

## 3.2 **Monte Carlo simulation**

To extend the zero temperature in first-principles calculations, we use Monte Carlo simulation (MCs), which is a probabilistic method based on five principles. Markov chain where the system evolution is temporal, ergodicity, i.e., the system can visit during its evolution all possible states, detailed balance which describes the system equilibrium, acceptance rate, and importance sampling, i.e., the considered states are those having a significant Boltzmann factor (p=e$^{-\Delta E/K_b T}$), where $\Delta E$ is the difference of energy between two configurations, K$_b$ is the Boltzmann constant, and T is the absolute temperature (Kelvin).

Based on the obtained results from ab initio calculations, we use (MCs) in order to explore the physical properties at finite temperature. We first extract the structural properties, as well as the polarization values, for the ground state using DFT calculations. We then propose a modified Heisenberg model (equation 4) to describe the ferroelectricity in a doped ferroelectric perovskite. This model includes the dipole interactions between the different ions as well as the substitution of bismuth by dysprosium into NBT matrix with a layered A-site configuration. This is used for the study of the transition temperature and hysteresis loops together with the temperature and doping effect on the electrical properties of these systems.

$$H = \left(-J_1 \sum_{i,j} \boldsymbol{p}_i^{Na} \cdot \boldsymbol{p}_j^{Na} - J_2 \sum_{k,l} [(1-c_k)\boldsymbol{p}_k^{Bi} + c_k \boldsymbol{p}_k^{Dy}] \cdot [(1-c_l)\boldsymbol{p}_l^{Bi} + c_l \boldsymbol{p}_l^{Dy}] - J_3 \sum_{m,n} [(1-c_m)\boldsymbol{p}_m^{Bi} + c_m \boldsymbol{p}_m^{Dy}] \cdot \boldsymbol{p}_n^{Na}\right) - \boldsymbol{E}\left(\sum_l [(1-c_l)\boldsymbol{p}_l^{Bi} + c_l \boldsymbol{p}_l^{Dy}] + \sum_i \boldsymbol{p}_i^{Na}\right) \quad (4)$$

Where $c_i$ is the occupation number of dysprosium sites, and $1-c_i$ is the bismuth occupation, E is the applied electric field, $J_1=0.29$ eV C/m², $J_2=0.066$ eV C/m², $J_3=0.61$ eV C/m² are the dipole interactions between electric dipoles **p** for different Na/Bi atoms computed using DFT data. The concentration of dysprosium and bismuth in an i-site is given by $<c_i>$ and $<1-c_i>$, respectively. We consider that these dipoles have a starting 111 orientation and can change their orientations as a function of temperature. According to the Metropolis algorithm, when a dipole tries to change its orientation, the process will choose a different dipole orientation arbitrarily with a probability that is proportional to the Boltzmann factor. Therefore, the change of orientations is either rejected or accepted, starting from different initial conditions [53]. Results were obtained using $10^5$ Monte Carlo steps per dipole with the discard of $10^4$ steps per site for equilibrium. Considering periodic boundary conditions with a total number of sites $N = L^3$ with $L = 4$. The thermodynamic quantities are calculated by means of the Metropolis algorithm. Therefore, the total polarization is given by:

$$\boldsymbol{P} = \frac{1}{N}\sum_{i=1}^{N} \boldsymbol{p_i} \qquad (5)$$

As mentioned before, experimental studies look forward to discovering novel NBT based systems that have improved properties required for practical applications. We focused in this section on the effect of dysprosium introduction together with the effect of temperature on the electrical properties of NBT.

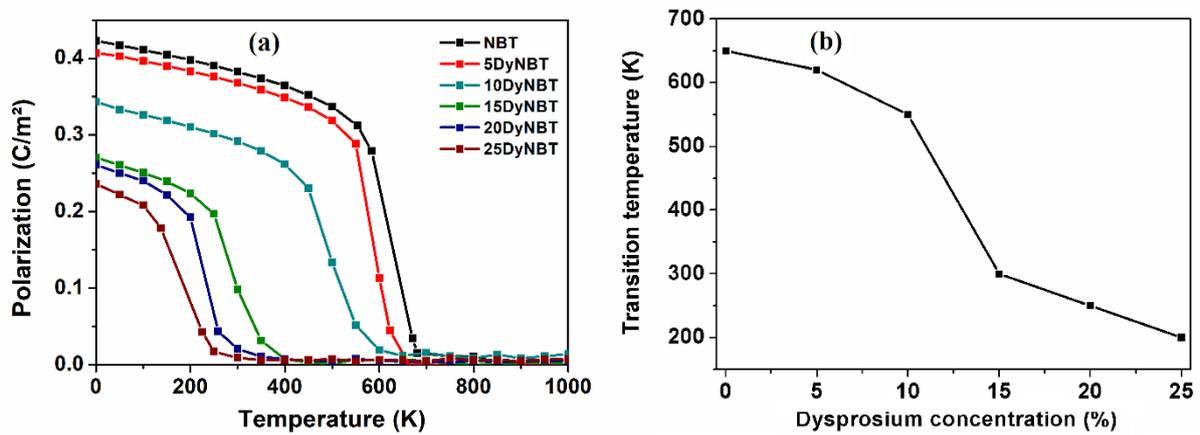

**Figure 6: (a)** Polarization as a function of temperature for different dysprosium concentrations, **(b)** transition temperature as a function of dysprosium concentrations.

Figure 6 (a) presents the polarization versus temperature for different dysprosium doped NBT systems. For the pure NBT compound (x=0), the transition temperature was observed to be of second-order around 650K, which is in good agreement with the value determined experimentally [10]. The introduction of dysprosium into NBT matrix seemed to decrease both the polarization value and the transition temperature from 0.42 C/cm² to 0.23 C/cm² and from 650K to 200K respectively, with the increase of x from 0% to 25% (figure 6 (a and b)). Noting that, starting from 15% of doping, the ferroelectricity of the system is found under room temperature (<300K) which agrees well with our recent experimental study in which a weak ferroelectric behavior (paraelectric like) was seen for concentrations x >10% [5].

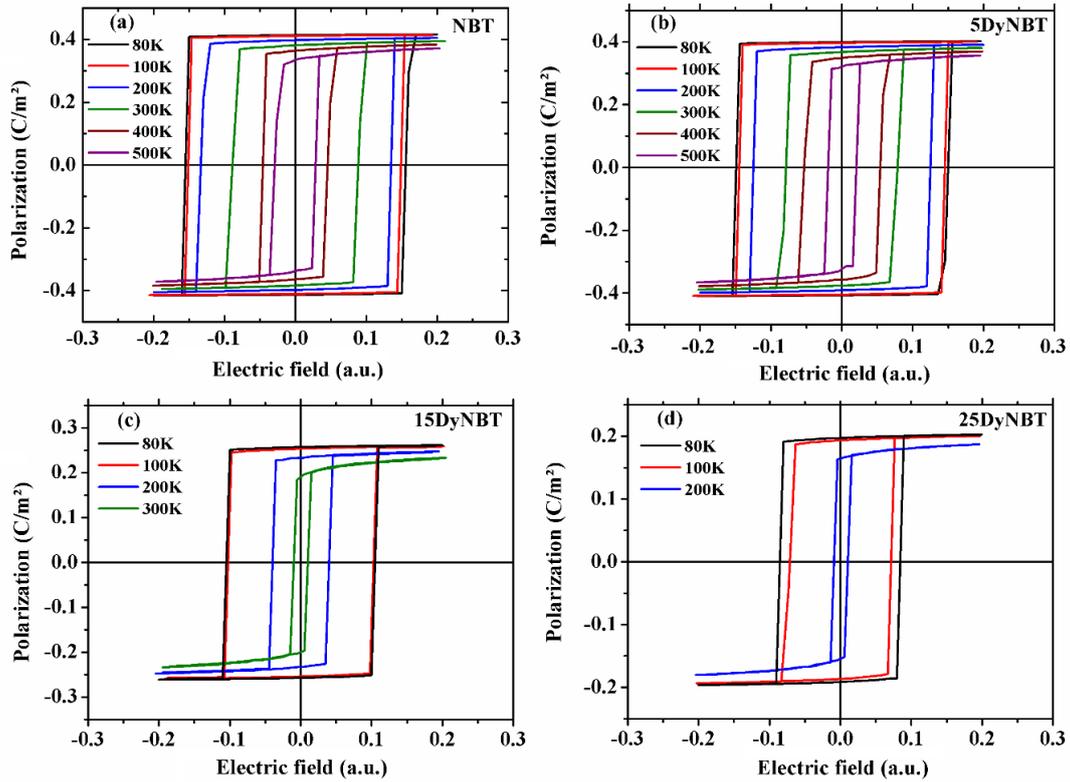

**Figure 7:** Polarization as a function of the electric field for different dysprosium concentrations.

In order to probe the response of our systems to an applied electric field, the ferroelectric hysteresis loops were computed. Moreover, we could also see the effect of the high-temperature application on the dipoles of our compounds. Figure 7 (a – d) present the hysteresis loops of some xDyNBT compositions (x = 0, 5, 15, 25%). We considered in figure 7 different temperatures starting from 80K to 500K. We can visualize a reduction in the area of hysteresis loops with an increase in temperature. Therefore, the decrease of the remanent polarization, as well as the coercive field, is due to the non-easy alignment of dipoles in the electric field direction owing to the temperature effect. In order to check out the effect of doping and temperature on the ferroelectric properties of NBT, we presented in table 3 values of the remanent polarization for different concentrations and at different temperatures (80K, 300K and 500K). The empty boxes in table 3 represent a paraelectric state for the given temperature.

We notice that the substitution of dysprosium in NBT matrix leads to a decrease in the polarization value. This behavior agrees well with the result found in DFT calculations.

**Table 3:** Remanent polarization for xDyNBT with x = 0 – 25% at different temperatures.

|       | $P_r$ (µC/cm²) | | | | | |
|-------|------|-------|--------|--------|--------|--------|
| T (K) | x=0  | x=5%  | x=10%  | x=15%  | x=20%  | x=25%  |
| 80    | 41.3 | 39.88 | 31.91  | 25.71  | 24.67  | 19.69  |
| 300   | 38   | 36.77 | 27.91  | 19.07  | -      | -      |
| 500   | 33.4 | 32.06 | -      | -      | -      | -      |

Notice that below the temperature transition, typical ferroelectric behaviors with square hysteresis loops are observed for all compositions (figure 7 (a – d)). Despite the fact that in experimental work, an antiferroelectric like behavior pinched hysteresis loop has been observed in 5DyNBT ceramics at high temperatures. This last antiferroelectric like behavior was found to be due to a reorientation of the polar vector in structure [54]. In fact, the R3c room-temperature structure of NBT has cations oriented along the $[111]_p$ direction, and with increasing temperature, the ferroelectric domains begin to disappear and are being replaced by Pnma orthorhombic sheets in which the cations are oriented along $[u0w]_p$ [55]. To obtain the AFE like behavior, we should take into account the effect of domains where each domain has a different dipolar orientation. This last effect has been described by Milhazes *et al.* for $BA_xBP_{1-x}$ system and by Misirlioglu *et al.* for $PbZrO_3$ system [56]. This study is being developed and results will be published elsewhere.

## 4 Conclusion

Using DFT calculations, we investigated the chemical ordering, electrical, optical, and magnetic properties of NBT and 25% dysprosium doped $Na_{0.5}Bi_{0.5}TiO_3$. We found that the most stable structure is given for the layered A-site configuration. While an A-site disorder can be perceived for DyNBT system. The emergence of a significant magnetic moment of 5$\mu_B$ was found for the doped system in contrast to the non-magnetic NBT. The different bonding interactions of bismuth and dysprosium with oxygen atoms seem to be the reason for the weakening of the polarization value in xDyNBT system. Monte Carlo simulation was used to study the transition temperature for different dysprosium concentrations (0 – 25%) together with the effect of temperature on the electric polarization. The remanent polarization and the coercive field was seen to decrease as a function of doping. We believe that our proposed model may be used for other doped ferroelectric perovskite compounds.

## Acknowledgments

Financial support by the Haute France Region/ FEDER (project MASEN) and H2020-RISE-ENGIMA-778072 project are gratefully acknowledged.